\documentclass[12pt]{article}%
\usepackage[cp1251]{inputenc}
\usepackage[english]{babel}
\usepackage{epsfig}
\usepackage{amssymb}
\usepackage{amsmath}%
\usepackage{amsfonts}%
\usepackage{graphicx}
\providecommand{\U}[1]{\protect\rule{.1in}{.1in}}
 0
\begin{document}

\title{Quantum corrections to static solutions of phi-in-quadro and Sin-Gordon models
via generalized zeta-function }
\author{Anatolij Zaitsev*, Sergey Leble\\Gdansk University of Technology, \\Faculty of Applied Physics and Mathematics, \\{\small ul. Narutowicza 11/12, 80-952 Gdansk, Poland,}\\Kant State University, {\small Faculty of Physics, }\\{\small Al.Nevsky st. 14, 236041, Kaliningrad, Russia. }\\{\small leble@mif.pg.gda.pl }\\
\\[2ex] }
\maketitle

\begin{abstract}
A general algebraic method of quantum corrections evaluation is  presented.
Quantum corrections to a few classical solutions (kinks and periodic) of
Ginzburg-Landau (phi-in-quadro) and Sin-Gordon models are calculated in
arbitrary dimensions. The Green function for heat equation with a soliton
potential is constructed by means of Laplace transformation theory and Hermit
equation for the Green function transform. The generalized zeta-function is
used to evaluate the functional integral and quantum corrections to mass in
quasiclassical approximation.

\end{abstract}

\renewcommand{\abstractname}{\small }

\section{Introduction.}

\subsection{General remarks.}

We consider one-dimensional field theory, based on nonlinear Klein-Gordon
equations, arising, for example of Sine-Gordon (SG) case, in kink models for
crystal structure dislocations \cite{Br}. In \cite{HKSF} the authors study the
diffusion of kinks. The Ginzburg-Landau (GL) model is very popular in
different aspects of solid state physics, e.g. for magnetics \cite{Wint}.

\subsection{Feynmann quantization of a classical field.}

An attention to Feynmann quantization formalism of a classical field was
recently attracted in connection with a link to a SUSY quasiclasic
quantization condition \cite{CC} suggested in \cite{Ju}; see, however,
\cite{Pe}.

A class of nonlinear Klein-Gordon equations in the case of static
one-dimensional solutions is reduced to
\begin{equation}
\label{KG}\phi^{\prime\prime}- V^{\prime}(\phi)=0, \phi= \phi(x), x\in R.
\end{equation}
Suppose the potential $V(\phi)$ is twice continuously differentiable; it
guarantees existence and uniqueness of the equations correspondent to
(\ref{KG}) Cauchy problem solution. The first integral of (\ref{KG}) is given
by
\begin{equation}
\label{E}W = \frac{1}{2}(\phi^{\prime2}-V(\phi),
\end{equation}
where W is the integration constant. The equation (\ref{E}) is ordinary
first-order differential equation with separated variables. As the phase
method shows, solutions of this equations belong to the following families:
constant, periodic, separatrix and the so-called "passing" one \cite{TD}. In
the case of phi-in-quadro (GL, non-integrable) model, the "potential" is
\[
V_{gl}(\phi) = \frac{g}{4}(\phi^{2}-\frac{m^{2}}{g})^{2},
\]
while in the case of Sin-Gordon (SG) model it is
\begin{equation}
\label{Vsg}V_{sg}(\phi) = \frac{2m^{4}}{3g}(1+\\
cos(\frac{1}{m}\sqrt{\frac{3g}{2}}x)).
\end{equation}
We modified this last model to fit it with the first one at small values of
the constant g, namely
\[
V_{sg} = \frac{13m^{4}}{12g}+ V_{gl} + O(g^{2}).
\]

In the papers of V.Konoplich \cite{Ko} quantum corrections to a few classical
solutions by means of Riemann zeta-function are calculated in dimensions $d >
1$. Most interesting of them are the corrections to the kink - the separatrix
solution of the field $\phi^{4}$ (GL) model. The method of \cite{Ko} is rather
complicated and it is desired to simplify it, that was the main target of our
previous note \cite{LeZa4}. We applied the Darboux transformations technique
with some nontrivial details missed in \cite{Ko}. The suggested approach open
new possibilities; for example it allows to show the way to calculate the
quantum corrections to matrix models of similar structure
, Q-balls \cite{CE} and periodic solutions of the models. The last problem is
posed in the review \cite{TD}.

The approximate quantum corrections to the solutions of the equation
(\ref{KG}) are obtained via the Feynmann functional integral method evaluated
by the stationary phase analog \cite{Jun}. It gives the following relation
\begin{equation}
\label{S}\exp[-\frac{S_{qu}}{\hbar}]\simeq\frac{A}{\sqrt{D}},
\end{equation}
where $S_{qu}$ denotes quantum action, corresponding the potential $V(\phi)$,
$A$ - some quantity determined by the vacuum state at $V(\phi) = 0$, and $\det
D$ is the determinant of the operator
\begin{equation}
\label{D}D = - \partial^{2}_{x} - \Delta_{y} + V^{\prime\prime}(\phi(x)).
\end{equation}
The variable $y \in R^{d-1}$ stands for the transverse variables on which the
solution $\phi(x)$ does not depend. The operator D appears while the second
variational derivative of the quantum action functional (which enter the
Feynmann trajectory integral) is evaluated. For the vacuum action $S_{vac}$
the relation of the form (\ref{S}) is valid if $S_{qu}$ is changed to
$S_{vac}$ and $D$ is replaced by the "vacuum state" operator $D_{0} = -
\partial^{2}_{x} - \Delta_{y}$. Then, the quantum correction
\begin{equation}
\label{qucor}\Delta S_{qu} = S_{qu} - S_{vac},
\end{equation}
is obtained by the mentioned twice use of the formula (\ref{S}) as
\begin{equation}
\label{qucor1}\Delta S_{qu} = \frac{ \hbar}{2}\ln(\frac{\det{D}}{\det{D_{0}}%
}).
\end{equation}
Hence, the problem of determination of the quantum correction is reduced to
one of evaluation of the determinants $D$ and $D_{0}$ ratio. The methodic of
the evaluation will be presented in the following section.

\subsection{The generalized Riemann zeta-function and Green function of heat
equation.}

The generalized zeta-function appears in many problems of quantum mechanics
and quantum field theories which use the Lagrangian $\mathcal{L}%
=(\mathbf{\partial}\phi)^{2}/2  - V(\phi)$ and it is necessary to calculate a
Feynmann functional integral in the quasiclassical approximation.

The scheme is following. Let $\{\lambda_{n}\} = S$ be a set of all eigenvalues
of a linear operator $L$, then, logarithm of the operator determinant is
represented by the formal sum over this set
\begin{equation}
\label{lndet}\ln(\det L)= \sum_{\lambda_{n}\in S}\ln\lambda_{n},
\end{equation}
where the sum in the r.h.s. is formal one, S means the operator's spectrum.
The generalized Riemann zeta-function $\zeta_{L}(s)$ is  defined by the
equality
\begin{equation}
\label{zeta0}\zeta_{L}(s) = \sum_{\lambda_{n}\in S}\lambda_{n}^{-s}.
\end{equation}
This definition should be interpreted as analytic continuation to the complex
plane of $s$ from the half plane $Re s > \sigma$ in which the sum in
(\ref{zeta0}) converges. Differentiating the relation (\ref{zeta0}) with
respect to $s$ at the point $s=0$ yields
\begin{equation}
\label{lndet}\ln(\det L)= \zeta_{L}^{\prime}(0).
\end{equation}

The generalized zeta-function (\ref{zeta}) admits the representation via the
diagonal $g_{D}$ of a Green function of the operator $\partial_{t}+L$. Such
representation is obtained as follows.

Let $\mathbf{r} \in\mathbb{R^{d}}$ be the set of independent variables of the
operator $L$; particularly, the operator $D$ of (\ref{D}) depends on
$\mathbf{r}=(x,\mathbf{y}) \in\mathbb{R} \times\mathbb{R}^{d-1}$, then, at
$t>0$,
\begin{equation}
\label{g}(\partial_{t}+L)g(t,\mathbf{{r},{r_{0}})=\delta({r}-{r_{0}})}%
\end{equation}
and
\[
g(t,\mathbf{{r},{r_{0}}})=0, \quad t<0.
\]
There is a representation in terms of the formal sun
\begin{equation}
\label{gD}g_{D}(t,\mathbf{{r},{r_{0}}) =\sum_{n} \exp[-\lambda_{n} t] \psi
_{n}(r)\psi_{n}^{*}(r_{0}),}%
\end{equation}
where the normalized eigenfunctions $\psi_{n}(\mathbf{r)}$ correspond to
eigenvalues $\lambda_{n}$ of the operator $L$. Put $L\equiv D$. Let us
introduce the function
\begin{equation}
\label{gamma}\gamma_{D}(t) = \int d \mathbf{r} g_{D}(t,\mathbf{{r},{r}) =
\sum_{n} \exp[-\lambda_{n} t],}%
\end{equation}
that follows from (\ref{gD}) and normalization. The integration in
(\ref{gamma}) is performed either along the axis in the kink case or over the
period in a case of periodic solutions. The Mellin transformation of
(\ref{gamma}) yields in
\begin{equation}
\label{gammazeta}\zeta_{D}(s) = \frac{1}{\Gamma(s)}\int_{0}^{+\infty}%
t^{s-1}\gamma_{D}(t)dt,
\end{equation}
where the $\Gamma(s)$ is the Euler Gamma-function.

The generalized zeta-function, defined by the relations (\ref{gamma}%
,\ref{gammazeta}), will be referred as the zeta-function of the operator $D$.

From the relation (\ref{gamma}) for the function $\gamma_{D}(t)$ it follows an
important property of multiplicity: \textit{if the operator D is a sum of two
differential operators $D = D_{1} + D_{2}$, which depend on different
variables, the following equality holds}
\begin{equation}
\label{gg}\gamma_{D}(t)=\gamma_{D_{1}}(t)\gamma_{D_{2}}(t).
\end{equation}

We will need the value of the function $\gamma_{D}(t)$ for the vacuum state,
when the operator $D=D_{0} = -\Delta$ is equal to the d-dimensional Laplacian.
In this case the formal sum in the r.h.s of (\ref{gamma}) goes to
d-dimensional Poisson integral
\begin{equation}
\label{Poi}\gamma_{D_{0}}(t) = \frac{1}{(2\pi)^{d}}\int_{\mathbb{R^{d}}%
}d\mathbf{k} \exp(-|\mathbf{k}|^{2}t) = (4\pi t)^{-d/2}.
\end{equation}
More generally, for a constant potential $\nu$,
\begin{equation}
\label{nu}\gamma_{D_{\nu}}(t) = (4\pi)^{\frac{1-d}{2}}\frac{\Gamma
(s+\frac{1-d}{2})}{\Gamma(s)}\nu^{\frac{d-1-2s}{2}}.
\end{equation}

The basic relation (\ref{qucor1}) points to a necessity of evaluation of the
determinants of the operators
\begin{equation}
\label{det}D=D_{0}+u(x), \quad D_{0}=-\partial_{x}^{2}-\Delta_{y}+\lambda
\end{equation}
where $\lambda$ is a positive number and $x\in R$ is one of variables, while
$\mathbf{y}\in R_{d-1}$ is a set of other variables. The operator $\Delta_{y}
$ is the Laplace operator in d-1 dimensions, u(x) is one-dimensional potential
that is defined by the condition
\begin{equation}
\label{pot}V^{\prime\prime}(\phi_{0}(x)) = \lambda+u(x),
\end{equation}
where $\phi_{0}(x)$ is the classical static solution of the equation of motion.

A quantum correction to the action in one-loop approximation for the classical
solution $\phi_{(}x)$ is calculated via zeta-function by the formula
\begin{equation}
\label{qucorM }\Delta\epsilon= -\zeta^{\prime}_{D}(0)/2,
\end{equation}
where
\begin{equation}
\label{zetaM}\zeta_{D}(s) = M^{2s}\int_{0}^{\infty}\gamma(t)t^{s-1}%
dt/\Gamma(s);
\end{equation}
here $\Gamma(s)$ is the Euler gamma function and M is a mass scale.

The function $\gamma(t)$ in the Mellin integral (\ref{zetaM}) is expressed via
the Green functions difference  $G(x,y,t;x_{0},y_{0},t_{0})$ and
$G_{0}(x,y,t;x_{0},y_{0},t_{0})$  of $\partial_{t}+D$ and $\partial_{t} +
D_{0}$ in the following way; let
\[
g(x,t) = G(x,y,t;x_{0},y_{0},0) - G_{0}(x,y,t;x_{0},y_{0},0),
\]
due to the translational invariance along y of operators D and $D_{0}$ the
function g does not depend on y; the contraction of $G_{0}$ is necessary for
deleting of ultraviolet divergence.

\section{Static solutions of the SG and $\varphi^4$ model}

Starting with SG model we integrate the equation (\ref{KG}) with
 the potential (\ref{Vsg}) arriving at the first-order differential
 equation
 with the parameter W (\ref{E}).
 \begin{equation}\label{Wsg}
   (\varphi')^2 =
   \frac{4m^4}{3g}(1+cos(\frac{1}{m}\sqrt{\frac{3g}{2}}\varphi))+
   2W.
 \end{equation}
The solutions are restricted and hence have the direct physical
relevance, if
\begin{equation}\label{condW}
    -\frac{4m^4}{3g}\leq W \leq 0,
\end{equation}
that follows from phase plane analysis. If $W=0$ the nontrivial
solutions are interpreted as kink and antikink
\begin{equation}\label{kink}
    \varphi(x)=\pm\sqrt{\frac{2}{3g}}\arcsin{\tanh(mx)}(mod\Phi),\qquad
    \Phi=2m\pi \sqrt{\frac{2}{3g}},
\end{equation}
while at the interval
$$
  -\frac{4m^4}{3g}< W < 0,
$$
the solution of (\ref{Wsg}) yields a periodic function expressed via
elliptic Jacobi function. To find it one plug $\varphi = \pm 2m
\frac{2}{3g}\arcsin{z},$ then the equation (\ref{Wsg}) goes to
\begin{equation}\label{sJac}
(z')^2=m^2(1-z^2)(1+\frac{3gW}{4m^4}).
\end{equation}
The solution of (\ref{sJac}) at the interval (\ref{condW}) is given
by
\begin{equation}\label{z=sn}
    z=ksn(mx;k),
\end{equation}
where
\begin{equation}\label{k}
    k=\sqrt{1+\frac{3gW}{4m^4}}
\end{equation}
is the module of the elliptic function. Hence
\begin{equation}\label{W}
    W = \frac{4(k^2-1)m^4}{3g}.
\end{equation}
Finally
\begin{equation}\label{varphi}
   \varphi = \pm
2m \frac{2}{3g}\arcsin{ksn(mx;k)}\quad (mod\Phi).
\end{equation}
The class of restricted solutions contains also
\begin{equation}\label{0}
    \varphi=0 (mod\Phi); W=-\frac{3m^4}{3g}
\end{equation}
and
\begin{equation}\label{const}
    \varphi=\pm\pi m \sqrt{\frac{2}{3g}} (mod \Phi) ; W=0.
\end{equation}
Other static solutions are obtained by shifts
$$
x \rightarrow x+x_0, \quad \varphi\rightarrow \varphi + \Phi,
$$
that follows from Klein-Gordon equation invariance and SG equation
potential periodicity.

 In the case of static solutions of the $\varphi^4$ model the potential and
the Lagrangian are determined by the formulas
$$V(\varphi) = g\varphi^4/4 - m^2\varphi^2/2,$$
$$L =
-(\varphi')^2 - g\varphi^4/4 + m^2\varphi^2/2,$$
 therefore the
equation of motion has the form
\begin{equation}\label{eqmot}
    \varphi''(x) + m^2\varphi - g\varphi^3 = 0,
\end{equation}
that yields (\ref{E}) in the form
\begin{equation}\label{Wphi4}
(\varphi')^2 = \frac{g}{2}(\varphi^2- \frac{m^2}{g})^2 + 2W.
\end{equation}
Its restricted solutions, as it follows from phase plane analysis,
exist if
\begin{equation}\label{Wphi4cond}
   -\frac{m^2}{4g}\leq W \leq 0.
\end{equation}
The separatrix  (W=0) solution of (\ref{Wphi4}) is the kink/antikink
\begin{equation}\label{kink}
    \varphi_0 = \pm\sqrt{\frac{2}{g}}b\tanh(bx)), \qquad b=\frac{m}{\sqrt{2}}.
\end{equation}

While inside the interval (\ref{Wphi4cond}) the equation
(\ref{Wphi4}) is expressed in terms of the elliptic Jacobi sinus
\begin{equation}\label{phi4sn}
   \varphi=\pm\sqrt{\frac{2}{g}}kb sn(bx;k), \qquad
   b=\frac{m}{\sqrt{1+k^2}}, \quad 0<k<1.
\end{equation}
The constant W is given by
\begin{equation}\label{Wsn}
    W = - (\frac{1-k^2}{1+k^2})^2\frac{m^4}{4g}.
\end{equation}

The family of the restricted solutions contains also the constant
vacuum ones (W=0)
\begin{equation}\label{phi4const}
    \varphi = \pm\frac{m}{\sqrt{g}}.
\end{equation}

After the substitution of (\ref{kink}) into (\ref{pot}) we obtain
the following potential u(x) :
\begin{equation}\label{pokink}
    u(x) = -6b^2/ch^2(bx),
\end{equation}
with the meaning of the constant $  b = m/\sqrt{2}. $ As a result
the two-level reflectionless potential of one-dimensional
Schr{\accent "7F o}dinger equation $ -
\partial^2_x + u(x) $ appears. Eigenvalues  and the
normalized eigenfunctions of which are correspondingly (its
numeration is chosen from above to lowercase).
$$\lambda_1 = - b^2, \qquad \psi_1(x) = \sqrt{3b/2} \sinh(bx)/cosh^2(bx);$$
$$\lambda_2 = - 4b^2,\qquad \psi_2(x) = \sqrt{3b}/2\cosh(bx).$$

\section{The energy of classic static solutions of SG and $\phi^4$ models}

Let us evaluate the energy of the nontrivial static solutions of
both models via the  energy density definitions
\begin{equation}\label{Ekink}
e(x) = \int_{-\infty}^{\infty}((\phi')^2/2+V(\varphi))dx,
\end{equation}
for kinks and, in a case of periodic solutions,
\begin{equation}\label{Ekink}
E = 2\int_{0}^{l}e(x)dx,
\end{equation}
 the constant "l" is the period of
the solution.

For SG kink:
\begin{equation}\label{Ekink}
    E_k = \frac{16m^2}{g},
\end{equation}
for a periodic soliton
\begin{equation}\label{Eper}
     E_p = \frac{8m^2}{g}[(1-k^2)K+2E],
\end{equation}
where K(k),E(k) - complete elliptic Legendre integrals.

\medskip

\section{The generalized zeta-function as analytic function.}

\medskip
Let us consider the Laplace transform of the one-dimensional Green
function defined in the Sec.1 by the relations 
(\ref{g},\ref{gD})
\begin{equation}\label{Gpx}
    g_L(t,x,x_0) =
    \frac{1}{2\pi\imath}\int_l\hat{g}_L(p,x,x_0)e^{pt}.
\end{equation}
where the integration is performed along the contour in complex
p-plane that contain all the singularities of the integrand . The
integrand  function $\hat{g}_L(p,x,x_0)$ may be considered as  the
Green function of the spectral equation
\begin{equation}\label{L+p}
    (L+p)\hat{g}_L(p,x,x_0) = \delta(x-x_0).
\end{equation}
The corresponding transform of $\gamma(t)$ is denoted as
\begin{equation}\label{gammap}
   \hat{ \gamma}(p) = \int\hat{g}_L(p,x,x)dx,
\end{equation}
the inverse transform of (\ref{gammap}) looks as (\ref{Gpx}) with
the same contour of integration
\begin{equation}\label{gammat}
    \gamma_L(t) = \int_l \hat{ \gamma}(p)e^{pt}dp.
\end{equation}

The inverse of the formula (\ref{gammap}) may be considered as a
base for analytic continuations. It is known from the general
properties  Laplace transform of a distribution that are zero at
$t<0$ \cite{Suk04}, so that the functions (\ref{gammap}) and
$\hat{g}_L(p,x,x_0)$ are analytic function at $Re p
> \sigma_0$.

It also shows that no necessity to solve the equation for the Green
function (\ref{g}), it is enough to solve the spectral problem
(\ref{L+p})

In a spirit of Hermit approach, see, e.g. \cite{UB} , the function
$\hat{g}_L(p,x,x)$ is a solution of bilinear equation
\begin{equation}\label{Hermit}
    2GG'' - (G')^2 - 4(u(x)-p)G^2+1=0,\qquad G(p,x) = \hat{g}_L(p,x,x)
\end{equation}
which in a case of reflectionless and finite-gap solutions is solved
more effectively than (\ref{L+p}).

In the case of  the periodic solution of the phi-in-quadro model the
potential has a "cnoidal" form
\begin{equation}\label{cnphi4}
   u(x)= - 6k^2b^2cn^2(bx;k)+(5k^2 - 1)b^2.
\end{equation}
This potential differs from second Lam$\acute{e}$s equation
 potential by the
constant hence its spectrum is two-gap one.

It is possible to cover all necessary classes of solutions of both
models (cases A,B for $\phi^4$ and C,D for SG) via the universal
representation by means of polynomials  (in p) $P,Q$
\begin{equation}\label{PQ}
    G(p,x)=P(p,z)/2\sqrt{Q(p,z)},
\end{equation}
where
$$z = sech^2(bx)$$
for kinks A,C, and
$$
z=cn^2(bx;k)
$$
for the periodic B,D.

In the $\phi^4$  (A,B case)the function P(p) is linear in p
polynomial while Q(p) is one of the third order. Plugging (\ref{PQ})
yields
\begin{equation}\label{linkPQ}
    b^2(\rho(z)(2PP''-(P')^2)+\rho'(z)PP')-(p+u(z))P^2+Q=0,
\end{equation}
the primes denote derivatives with respect to z, while
\begin{eqnarray}
  \rho &=& \{\begin{array}{c}
     \\        z^2(1-z), cases \quad A,B; \\
               z(1-z)(1-k^2+k^2z), cases \quad C,D. \\
             \end{array}\\
  u(z) &=& \{\begin{array}{c}
               b^2(1-2z), case \quad (A) \\
               b^2(2k^2-1-2k^2z), case \quad (B) \\
               2b^2(2-3z), case \quad (C) \\
              b^2(5k^2-6k^2z), case \quad (D) \\
             \end{array}
\end{eqnarray}

Let us start with the case (A). The form of the polynomial is
determined from well-known facts of the reflectionless potentials
theory.
\begin{equation}\label{Q}
   Q=p^2(p+b^2).
\end{equation}
The case (B)
\begin{equation}\label{RQB}
    P=p+P_1(z), \quad Q=p^3+q_2p^2+q_1p+q_0m
\end{equation}
includes (A) in a sense that for the case $q_2=b^2,\quad q_1=0,
\quad q_0=0$.

Substituting (\ref{RQB}) into (\ref{linkPQ}) gives for each power of
$p = 0,1,2$
\begin{equation}\label{sys}
    \begin{array}{c}
       -2P_1(z)-u(z)+q_2=0, \\
       b^2(2\rho(z)P_1''+\rho'(z)P_1')-P_1^2 - 2u(z)P_1+q_1.\\
        b^2(\rho(z)(2P_1P_1''-P_1^2)+\rho'(z)P_1P_1'-u(z)P_1^2+q_0=0. \\
     \end{array}
\end{equation}
respectively. In the case (A) it gives
\begin{equation}\label{Arez}
    P_1=b^2z.
\end{equation}
For the case (B) we begin from
\begin{equation}\label{P1}
    P_1=k^2b^2z,
\end{equation}
that is the result of substitution of $P_1$ from the first equation
of (\ref{sys}) into the second one. Next, from the third equation
yields
\begin{equation}\label{Qrez}
    Q=p(p^2+(1-k^2)b^2)(p-k^2b^2),
\end{equation}
with the simple roots $p_i$.

Going to the cases (C,D), generally
\begin{equation}\label{PQD}
    P=p^2+P_1(z)p+P_2(z), \quad
    Q=p^5+q_4p^4+q_3p^3+q_2p^2+q_1p+q_0,
\end{equation}
while in the particular case of (C)
$$
q_0=0,\quad q_1=0,q_2=36b^6,\quad q_3=33b^4,q_5=10b^2.
$$
A substitution of (\ref{PQD}) into the Hermit equation splits in the
system
\begin{equation}\label{splitD}
    \begin{array}{c}
       -2P_1  - u  + q_2 = 0, \\
    -2P_2  - P_1^2  - 2 u P_1 +b^2(2\rho P_1"+\rho'P_1')+q_3,    \\
       b^2(\rho (2P_2''+2P_1P_1''-(P_1')^2)+\rho'(P_2'+P_1P_1'))-2P_1P_2)-u(2P_2+P_1^2)+q_2 =0 ,\\
       b^2(2\rho(2P_1''P_2+P_1'P_2'+P_1P_2'')+\rho'(P_1P_2')+P_1'P_2))-P_2^2 - 2uP_1P_2 + q_1,\\
       b^2(\rho(2P_2P_2''-P_2'^2))+\rho'P_2^2+q_0.\\
     \end{array}
\end{equation}
The arguments in (\ref{splitD}) are omitted. Solving the system
yields for the case D
\begin{equation}\label{P1P2}
P_1(z)=\alpha z + \beta, \qquad P_2(z)=az^2 +b z+c,
\end{equation}
where
\begin{equation}\label{alpha}
    \begin{array}{c}
      \alpha = 3k^2b^2, \\
      \beta = \frac{(1-5k^2)b^2 -q_4}{2},\\
      a = 9k^4b^4, \\
      b = \frac{3k^2b^2(q_4+(1-11k^2)b^2)}{2},  \\
      c = \frac{4q_3 -q_4^2+2(1-5k^2)b^2q_4+3(1-6k^2+21k^4)b^4}{8},\\
    \end{array}
\end{equation}
are functions only of $k,b$, where
\begin{equation}\label{q}
    \begin{array}{c}
       q_0 = 0, q_1= -27k^2(1-k^2)b^8, \\
        q_2=-9(1-3k^2-3k^4-k^6)b^6, q_3 = 3(1+k^2+k^4)b^4, q_4=5(1+k^2)b^2.  \\
      \end{array}
\end{equation}
 Finally,
 \begin{equation}\label{Q}
    Q=\prod_{i=1}^{i=5}(p-p_i),
\end{equation}
where the polynomial $Q$ the simple roots $p_i$ are ordered so that
(0<k<1)
\begin{equation}\label{p_i}
    - (2\sqrt{1-k^2+k^4}+1+k^2)b^2< -3b^2<-3k^2b^2<0<-
    (2\sqrt{1-k^2+k^4}-1-k^2)b^2.
\end{equation}

 Let us pick up the expressions determining
$\hat{\gamma}(p)$:
\begin{equation}\label{hatgammap}
    \hat{\gamma}(p) = \int\frac{P(z)}{3\sqrt{Q}}dx =
    \frac{p^2}{2\sqrt{Q}}\int dx+ \frac{p}{2\sqrt{Q}} \int(\alpha z+\beta) dx +
    \frac{1}{2\sqrt{Q}} \int(az^2 +b z+c) dx.
\end{equation}
So we need three integrals over the period $K$.
\begin{equation}\label{intz}
    \begin{array}{c}
       \int_0^K dx ,
       \int_0^K z dx  ,
\int_0^K z^2dx.
           \end{array}
\end{equation}
Let us go to the variable z, $ dz=d(cn^2(x))= - 2cn(x)cn(x)dn(x)dx =
2 sn(x)$, a bit more convenient to put $y=1-z, dy=-dz$
\begin{equation}\label{inty}
     \begin{array}{c}
        \int_0^K dx = \int_0^1\frac{dy}{y(1-y)(1-k^2y)} = 2K(k), \\
       \int_0^K sn^2(x;k)dx = \frac{1}2{}\int_0^1\frac{ydy}{y(1-y)(1-k^2y)} = \frac{K(k)-E(k)}{k^2}, \\
       \int_0^K sn^4(x;k) dy = \int_0^1\frac{z^2dz}{z(1-z)(1-k^2z)} = \frac{1}{3k^4}( (2+k^2)K(k)  - 2 (1+k^2)E(k)) . \\
     \end{array}
\end{equation}
hence
\begin{equation}\label{hatgammap'}
\begin{array}{c}
  \hat{\gamma}(p) = 2K(k)\frac{p^2}{2\sqrt{Q}} + \alpha (-\frac{K(k)-E(k)}{k^2}+ (\beta -\alpha)2K(k)\frac{p}{2\sqrt{Q}}+ \\
  (a \frac{(2+k^2)K(k)  - 2 (1+k^2)E(k)}{3k^4} + (b-2a) \frac{K(k)-E(k)}{k^2}+
    (a -b +c)2K(k)) \frac{1}{2\sqrt{Q}}.\\
\end{array}
   \end{equation}
or, finally
\begin{equation}\label{zeta}
    \zeta(s) = M^{2s}\int_0^\infty (\int_l \hat{ \gamma}(p)e^{pt}dp)t^{s-1}dt/\Gamma(s);
\end{equation}
which is expressed via integrals
\begin{equation}\label{intp}
  \begin{array}{c}
                                     \int_l \frac{e^{pt}}{\sqrt{Q}} dp \\
                                      \int_l \frac{pe^{pt} }{\sqrt{Q}} dp  \\
                                     \int_l \frac{p^2e^{pt}}{\sqrt{Q}} dp \\
                                  \end{array}
\end{equation}
by a contour that contains all branch points of the integrands.
\section{SG kinks}

The results of the previous sections allow to evaluate corrections to actions
for all four (A,B,C,D) cases. The results for $\phi^{2}$ model kinks are
well-known, see, e.g. \cite{LeZa4}, where a table for the dimensions d=1,2,3,4
is listed. These results fit the case B, which hence was strictly verified.

Let us present the formulas for the kinks of the SG model. The substitution of
the expressions from (\ref{PQ},\ref{Q},\ref{RQB},\ref{Arez}) yields a
divergence of the integral for (\ref{hatgammap}) $\hat{\gamma}(p)$. To
regularize the Green function let us divide itd Laplace transform into two
parts as
\begin{equation}
\label{Gck}G(p,x) = G_{c}(p) + G_{k}(p,x),
\end{equation}
where the part $G_{c}(p)$ is the Green function diagonal (the solution of
(\ref{Hermit}) for a constant potential:
\begin{equation}
\label{Gc}G_{c}=\frac{1}{2\sqrt{p+b^{2}}}.
\end{equation}
The kink part is easily constructed via (\ref{PQ},\ref{Q},\ref{RQB}%
,\ref{Arez}):
\begin{equation}
\label{Gk}G_{k}=\frac{b^{2}sech^{2}(bx)}{2p\sqrt{p+b^{2}}}.
\end{equation}
The representation (\ref{Gck}) results in two contributions of the quantum
corrections to the kink (antikink) mass.

The first one coinsides with (\ref{nu}) for $\nu=m^{2}$.
\begin{equation}
\label{nu}\gamma_{D_{0}}(t) = (4\pi)^{\frac{1-d}{2}}\frac{\Gamma(s+\frac
{1-d}{2})}{\Gamma(s)}m^{ d-1-2s },
\end{equation}
and the second one is obtained by the general scheme, namely
\begin{equation}
\label{gammak}\gamma_{k}(p) = \int_{-\infty}^{\infty}G_{k}(p,x)dx.
\end{equation}
Plugging (\ref{Gk}) into (\ref{gammak}) yields
\begin{equation}
\label{gamkp}\gamma_{k}(p)=\frac{b}{p\sqrt{p+k^{2}}}.
\end{equation}
The value of the integral $\int_{-\infty}^{\infty}sech^{2}(bx) = 2/b$ is taken
into account. the corresponding function (\ref{gamma}) or, more exactly
(\ref{gammat}) will be denoted $\gamma_{L}^{R}$, the index R label the result
of a renormalization provided by the division (\ref{Gck}). The function may be
found directly from a table (\cite{BE}) but we would explain the result as an
example for further development of the renormalization procedure. Note that
the cut for the radical $\sqrt{p+b^{2}}$ is made along the Re p -axis from $-
\infty$ to $-b^{2}$, the branch with $\sqrt{p+b^{2}} = i\sqrt{|p+b^{2}|}$ on
the upper and $\sqrt{p+b^{2}}=-i\sqrt{|p+b^{2}|}$ on the lower bounds of the
cut is chosen. After the deformation of the integral contour one has for $t >
0$
\begin{equation}
\label{gamRt}\gamma_{L}^{R} = 1 - \frac{b}{\pi}\int_{0}^{\infty}\frac{\exp
(\xi+ b^{2})t}{(\xi+b^{2})\sqrt{\xi}}d \xi.
\end{equation}
The formula (\ref{gamRt}) gives an expansion of the resolvent of the operator
$\partial_{t}+L$ by the operator $L$ spectrum. The direct substitution of
(\ref{gamRt}) into (\ref{gammazeta}) leads to a divergent integrals. However
in this case a renormalization is not necessary, the integral in the r.h.s. of
(\ref{gamRt}) is expressed in terms of the error function
\[
Erf(z) = \frac{2}{\sqrt{\pi}} \int_{0}^{z} e^{-\tau^{2}}d\tau.
\]
namely, after some change of variables
\begin{equation}
\label{gamkt}\gamma_{k}(t) = Erf(b\sqrt{t}) = \frac{2b\sqrt{t}}{\sqrt{\pi}%
}\int_{0}^{1}\exp[-b^{2}t\tau^{2}]d\tau.
\end{equation}
The same result gives \cite{BE}. The Mellin transform of this representation
gives the following expression of zeta function of the operator L.
\begin{equation}
\label{zetaL}\zeta_{L}^{R}(s) = \frac{2b^{-2s}}{\sqrt{\pi}}\frac
{\Gamma(s+1/2)}{\Gamma(s)}\int_{0}^{1}\tau^{-2s-1}d\tau= -\frac{b^{-2s}}%
{\sqrt{\pi}}\frac{\Gamma(s+1/2)}{\Gamma(s+1)}.
\end{equation}
The integral in (\ref{zetaL}) converged only at $Re s <0$ but gives the
analytical continuation for all $s \in C$ excluding the poles of $\Gamma(s+1)$.

Substituting the result (\ref{zetaL}) into one arising from (\ref{gg}) with
account of (\ref{nu}) yields
\begin{equation}
\zeta_{L}^{D}(s)=-4(4\pi)^{\frac{d}{2}}m^{d-1-2s}\frac{\Gamma(s+1-d/2)}%
{(2s+1-d)\Gamma(s)}.\label{zetaLf}%
\end{equation}
The condition $b=m$ is taken into account. Differentiation of
(\ref{zetaLf}) by $s$ at the point $s=0$ give the desired
correction (\ref{qucorM })
\begin{equation}
\begin{array}{c}
  -\frac{1}{2}\frac{d\left(  \zeta_{L}^{D}(s)\right)  }{ds}= \\
  -2(4\pi)^{\frac{d}{2}}%
m^{d-2s-1}\frac{\Gamma\left(  s-\frac{1}{2}d+1\right)
}{\Gamma\left( s\right)  \left(  2s-d+1\right)  ^{2}}\left(
\left(  d-2s-1\right)  \left(
\operatorname{Psi}\left(  -\frac{1}{2}d+s+1\right)  -2\ln m-\operatorname{Psi}%
\left(  s\right)  \right)  \ +2\right)  \label{final}
\end{array}
\end{equation}
next we plot the dependence of the correction $\frac{d\left( \zeta_{L}%
^{D}(0)\right) }{ds}$ on m (Fig \ref{z}).%
\begin{center}
\begin{figure}
  \includegraphics[width=4in]{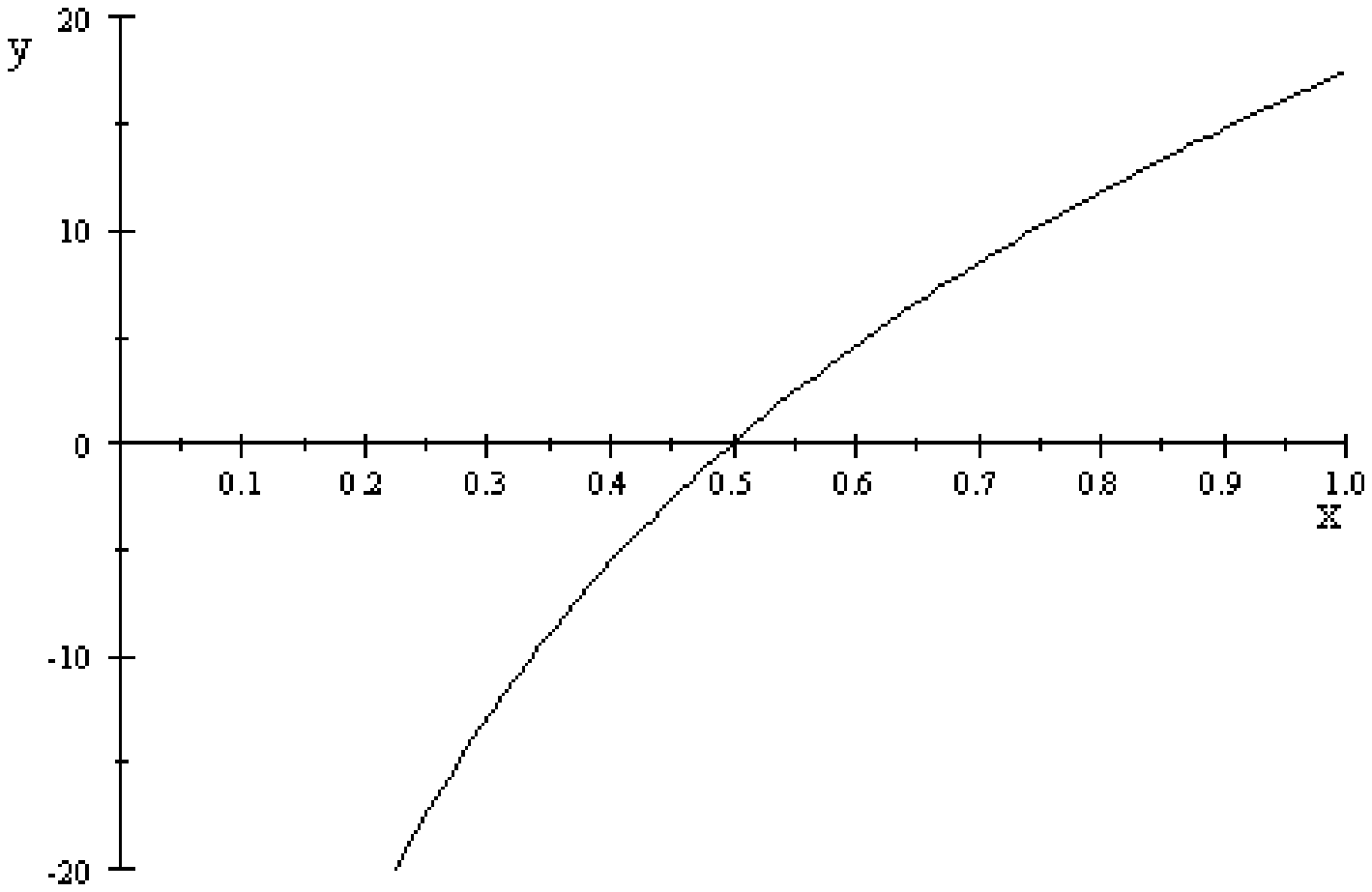}\\
      \caption{d=1, dependence of the correction $\Delta \epsilon$  to mass on the parameter m}\label{z}
\end{figure}
\end{center}
\begin{center}
\begin{figure}
  \includegraphics[width=4in]{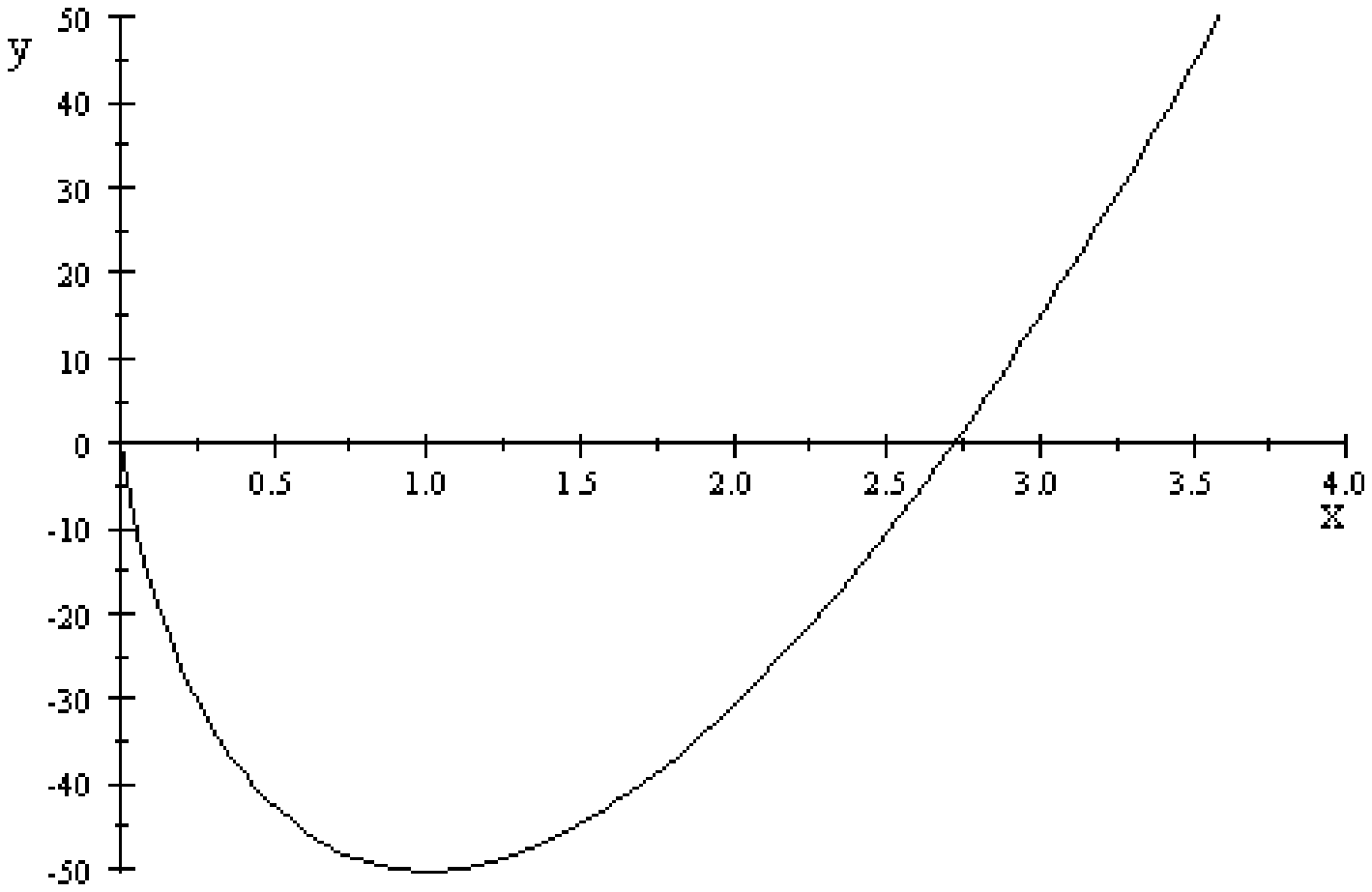}\\
      \caption{d=2, dependence of the correction $\Delta \epsilon$  to mass on the parameter m}\label{z}
\end{figure}
\end{center}
\begin{center}
\begin{figure}
  \includegraphics[width=4in]{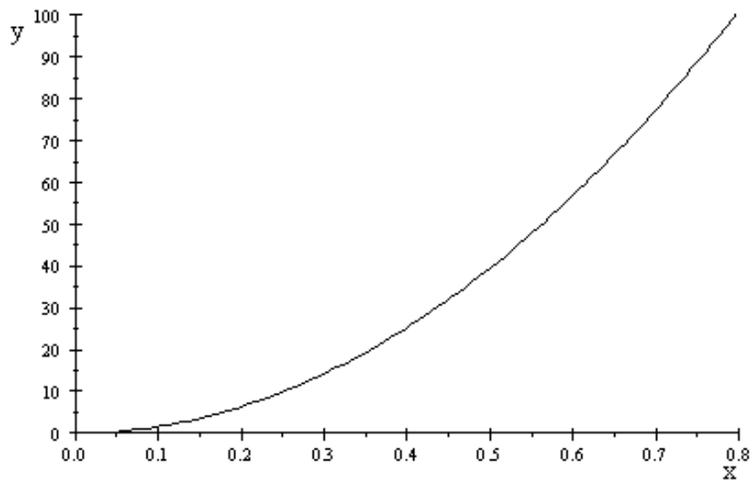}\\
      \caption{d=3, dependence of the correction $\Delta \epsilon$  to mass on the parameter m}\label{z}
\end{figure}
\end{center}
We would remind about the choice of the constant  g=2 in the last
sections.

\section{Conclusion}

In the next paper we admit that the potential u(x) from (\ref{pot}) has the
form of n-level reflectionless potential
\begin{equation}
\label{nlevel}u(x) = -n(n+1)b^{2}/\cosh^{2}(bx).
\end{equation}
with eigenvalues $\lambda_{m} = -m^{2}b^{2}$, $m=1,...,n$. We would note for
further generalization means that this function may be considered as
degenerate limit of n-gap Lame potential of Hill equation. Such potentials
correspond to higher solitonic models.

The kink case corresponds $n = 2, \lambda= n^{2}b^{2}$. The quantum correction
to its action will be calculated in sec.4 from general formulas of sec.3.

Let us note that all results related to the scalar nonlinear Klein-Gordon
equation may be applied directly to many-component model $\phi(x)\in
\mathbb{S^{m}}$ with account of $\mathbb{SO}(m)$ symmetry. The equation
(\ref{KG}) takes the form
\begin{equation}
\label{KGm}- \phi^{\prime\prime}+ V^{\prime}(||\phi||)\frac{\phi}{||\phi||} =
0.
\end{equation}
The scalar operator D is defined by (\ref{D}) goes to the matrix one
\begin{equation}
\label{Dm}D = [- \partial^{2}_{x} - \Delta_{y} + V^{\prime}(||\phi
||)\frac{\phi}{||\phi||}]I_{m} + [(V"(||\phi||)) - V^{\prime}(||\phi
||)\frac{\phi\otimes\phi}{||\phi||^{3}}].
\end{equation}
The technique developed in this paper is transported to quantum corrections
for periodic static solutions of $\phi^{4}$ model:
\begin{equation}
\label{24}\phi_{0}(x) = \frac{km}{1+k^{2}}\sqrt{\frac{2}{g}}sn(\frac{mx}%
{\sqrt{1+k^{2}}};k),\quad0< k \leq1;
\end{equation}
where $k$ is a module of the elliptic function. When $k=1$ the  formula
(\ref{24}) goes to one for kink (\ref{kink}). The  substitution of (\ref{24})
into (\ref{pot}) yields in two-gap Lame potential that is embedded in Darboux
Transformations theory by chain representation \cite{BL} that give a
possibility to derive the Green function analogue for this case. The results
will be published elsewhere. Some recent papers open new field for
applications \cite{B,MB}.

\end{document}